\documentclass[manuscript]{aastex} 
\usepackage{graphicx,epsfig}
\usepackage{natbib}                              
\usepackage{rotating}

\usepackage{txfonts}
\usepackage{longtable,lscape,url}
\usepackage{rotating}

\shorttitle{Characterization of a sample of intermediate-type AGN II.}
\shortauthors{Ben\'itez et al.}

\begin{document}

\title{CHARACTERIZATION OF A SAMPLE OF INTERMEDIATE-TYPE AGN. II. Host Bulge Properties and Black Hole Mass Estimates}
\shorttitle{}
\author{Erika Ben\'itez\altaffilmark{1}, Jairo M\'endez-Abreu\altaffilmark{2,3}, Isaura Fuentes-Carrera\altaffilmark{4}, Irene Cruz-Gonz\'alez\altaffilmark{1}, 
Benoni Mart\'inez\altaffilmark{1}, Luis L\'opez-Martin\altaffilmark{2,3}, Elena Jim\'enez-Bail\'on\altaffilmark{1}, Vahram Chavushyan\altaffilmark{5}, Jonathan Le\'on-Tavares\altaffilmark{6}}

\altaffiltext{1}{Instituto de Astronom\'ia, Universidad Nacional Aut\'onoma de M\'exico, Apdo. Postal 70-264, M\'exico D.F. 04510, M\'exico}

\altaffiltext{2}{Instituto de Astrof\'{\i}sica de Canarias, 38200 La Laguna, Tenerife, Spain}

\altaffiltext{3}{Departamento de Astrof\'isica, Universidad de La Laguna, E-38205 La Laguna, Tenerife, Spain}

\altaffiltext{4}{Escuela Superior de F\'isica y Matem\'aticas, Instituto Polit\'ecnico Nacional (ESFM-IPN), U.P. Adolfo L\'opez Mateos, M\'exico D.F. 07730, M\'exico}

\altaffiltext{5}{Instituto Nacional de Astrof\'{\i}sica, \'Optica y Electr\'onica, Apdo. Postal 51-216, 72000 Puebla, M\'exico}

\altaffiltext{6}{Aalto University Mets\"ahovi Radio Observatory, Mets\"ahovintie 114, 02540, Kylm\"al\"a, Finland}

\email{erika@astro.unam.mx}

\accepted{December 4 2012}

\begin{abstract}
We present a study of the host bulge properties and their relations with the black hole mass on a sample of 10 intermediate-type active galactic nuclei (AGN).  Our sample consists mainly of early type spirals, four of them hosting a bar. For  {\bf 70$^{+10}_{-17}\%$} of the galaxies we have been able to determine the type of the bulge, and find that these objects probably harbor a pseudobulge or a combination of classical bulge/ pseudobulge, suggesting that pseudobulges might be frequent in intermediate-type AGN. In our sample, $50\pm14\%$ of the objects show double-peaked emission lines. Therefore, narrow double-peaked emission lines seem to be frequent in galaxies harboring a pseudobulge or a combination of classical bulge/ pseudobulge. Depending on the bulge type, we estimated the black hole mass using the corresponding $M_{BH} - {\sigma*}$ relation and found them with a range of: {\bf 5.69$\pm$0.21 $<$ $\log M_{BH}^{\sigma*}$ $<$ 8.09$\pm$0.24. Comparing these $M_{BH}^{\sigma*}$  values with masses derived from {\bf the FWHM of H$\beta$ and the continuum luminosity at 5100 \AA \ from their SDSS-DR7 spectra ($M_{BH}$}) we find that eight out of ten ({\bf 80$^{+7}_{-17}$\%})  galaxies have black hole masses that are compatible within a factor of 3.  This result would support that $M_{BH}$ and $M_{BH}^{\sigma*}$ are the same for intermediate-type AGN as has been found for type 1 AGN. However, when the type of the bulge is taken into account only 3 out of the 7 (43$^{+18}_{-15}\%$) objects of the sample have their $M_{BH}^{\sigma*}$ and $M_{BH}$ compatible within 3-$\sigma$ errors. We also find that estimations based on the $M_{BH}-\sigma*$ relation for pseudobulges are not compatible in 50$\pm20\%$ of the objects.
}

\end{abstract}

\keywords{galaxies: active -galaxies: bulges - galaxies: photometry}




\section{Introduction}\label{introduction}

The  connection between black  hole  mass  ($M_{\rm  BH}$) and  several
properties of the host galaxy bulges have been a remarkable finding and
have been extensively  used in large samples of  active galactic nuclei
\citep[AGN; see][for a review]{2005SSRv..116..523F}.  Among them, the
so       called       $M_{\rm       BH}-\sigma_{\star}$       relation
\citep{2000ApJ...539L...9F,2000ApJ...539L..13G},   which  relates  the
$M_{\rm  BH}$ with  the  central stellar  velocity  dispersion of  the
galaxy ($\sigma_{\star}$), has been  demonstrated to be a reliable way
to estimate $M_{\rm BH}$.   However, a recent work by \citet[][see
  also \citealt{2008ApJ...688..159G}]{2008MNRAS.386.2242H} showed that
the  nature of  the host  galaxy  bulge must  be carefully  considered
before attempting any estimation of  the $M_{\rm BH}$.  This is mainly
due to  the existence of a  dichotomy in the  properties and formation
mechanisms of bulges \citep{1997ARA&A..35..637W,2004ARA&A..42..603K}.

On one  hand, the merger of  small galaxies has been  suggested as the
main  path for bulge  formation \citep{1993MNRAS.264..201K},  which is
supported by  the homogeneous bulge  stellar populations of  the Milky
Way and  M31 \citep{2003A&A...399..931Z, 2003AJ....125.2473S}.  Bulges
formed in such mergers are  termed {\it classical} bulges (CB) and are
similar  to low-luminosity  ellipticals. Their  light  distribution is
described         by        a        de         Vaucouleurs        law
\citep[e.g.,][]{1995MNRAS.275..874A}.  They  are composed primarily by
old                 Population                 II                stars
\citep{2003A&A...407..423M,2005ApJ...621..673T}, which have spheroidal
or   weakly    triaxial   distributions   \citep{2010A&A...521A..71M}.
Alternatively,  bulges   may  form  via   internal  secular  processes
\citep[see][for a review]{2004ARA&A..42..603K}  such as bar-driven gas
inflows, bending instabilities,  and clump instabilities. Evidence for
secular   bulge    ({\it   pseudobulge})   formation    includes   the
near-exponential            bulge            light            profiles
\citep{1994MNRAS.267..283A,2008AJ....136..773F}. Pseudobulges (PB) are
dominated  by  Population  I stars  \citep{2006MNRAS.366..510T},  they
appear to have lower metallicity \citep{2007MNRAS.380..506G} and lower
$\alpha$/Fe enhancement  with respect to the big  bulges of early-type
galaxies   \citep{2008MNRAS.389..341M}.   They   also  show   a  clear
correlation      between     bulge     and      disk     scale-lengths
\citep[e.g.,][]{2008A&A...478..353M},       substantial       rotation
\citep{2004ARA&A..42..603K},  and the  presence of  boxy-peanut shaped
bulges      \citep[e.g.,][]{2008ApJ...679L..73M}.     In     addition,
\citet{2007MNRAS.381..401L}     have     demonstrated    that
pseudobulges are  more prevalent among normal  galaxies than classical
bulges.

As previously mentioned, a common  way to estimate the $M_{\rm BH}$ is
through  the  $M_{\rm  BH}$-$\sigma_{\star}$ relation:  $\log  (M_{\rm
  BH}/M_{\odot})\,=\,\alpha\,+\,\beta\,      \log(\sigma_{\star}/200)$.
However,  considering  that  pseudobulges  have  properties  that  lie
between those  of classical  bulges and those  of disks,  the question
arises whether  $M_{\rm BH}$ masses should  correlate differently with
pseudobulge    velocity    dispersions.     \citet{2008MNRAS.386.2242H}
investigated  the  $M_{\rm  BH}$-$\sigma_{\star}$  relation  for  disk
galaxies  and found  that  it  is different  for  pseudobulges over  a
3$\sigma$ significance  level. Moreover,  he noted that  pseudobulges host
smaller black holes than classical  bulges do.  One of his conclusions
is that black holes (BH) form earlier than their host pseudobulges and that
their growth  is insignificant once  the pseudobulges are  formed.  He
also  suggested  that  AGN   fueling  is  less  efficient  in  secular
processes.      Similar     results     were     also     found     by
\citet{2008ApJ...688..159G},      \citet{2008ApJ...680..143G},     and
\citet{2009ApJ...698..812G}.    However,  other  works   performed  by
\citet{2009ApJ...692..856B}  and  \citet{2009ApJ...698..198G} did  not
find these  differences. Recently, \citet{2011Natur.469..374K} claimed
that nor pseudobulge luminosity, {\bf nor velocity dispersions of
  their hosts galaxies} correlate with $M_{\rm BH}$ mass
since  secular evolution  processes  lead to  no  coevolution of  both
components. Therefore, it is clear that when studying the $M_{\rm BH}$
it is important  to know whether the central  component is a classical
bulge, a pseudobulge, or a mix of both.

In  this work,  we  present new  observations of  a sample  of 10
  intermediate-type  AGN using the  Nordic Optical  Telescope (NOT).
We performed  a careful morphological analysis of  the sample, derived
the surface-brightness profile for each galaxy and obtained their main
structural parameters.  In particular,  we derived the S\'ersic index,
which along  with other criteria, we  used to establish  the nature of
each bulge, and  then estimated the $M_{\rm BH}$  for all objects using
different  correlations  according  to the  nature of  the bulge.
These mass  estimates  were  also  compared  with {\bf BH masses, 
$M_{\rm  BH}$ derived in Ben\'\i tez et al. (2012, hereafter Paper I) 
using SDSS-DR7 spectra and the relation given by \citet{2006ApJ...641..689V}}.
Differences  and similitudes were  analysed in  the light of  the nature 
of  the bulge. In addition, in Paper I narrow double-peaked AGN were 
detected in five galaxies of  the sample. This result is used  in order 
to compare the nature of the bulges with  the presence  of single-  
or double-peaked  emission  lines in intermediate-type AGN.

The  paper  is  organized as  follows:  Sect.~\ref{observations}
 describes the observations and data  reduction process. The  host galaxy
 photometric decomposition  and the galaxy  structural parameters are
 explained  in  Sect.~\ref{photdec}.  The  results  obtained for  the
 complete  sample  and  for  the  individual  objects  are  shown  in
 Sect.~\ref{results}.   The discussion and  conclusions are  given in
 Sect.~\ref{discussion}.   The  cosmology adopted  in  this work  is
$H_{0}=$70    km    s$^{-1}$    Mpc$^{-1}$,    $\Omega_{m}=$0.3    and
$\Omega_{\lambda}=$0.7.

\vskip.2cm

\section{Observations and Data Reduction}
\label{observations}

\begin{figure*}
{\includegraphics[angle=0,width=\textwidth]{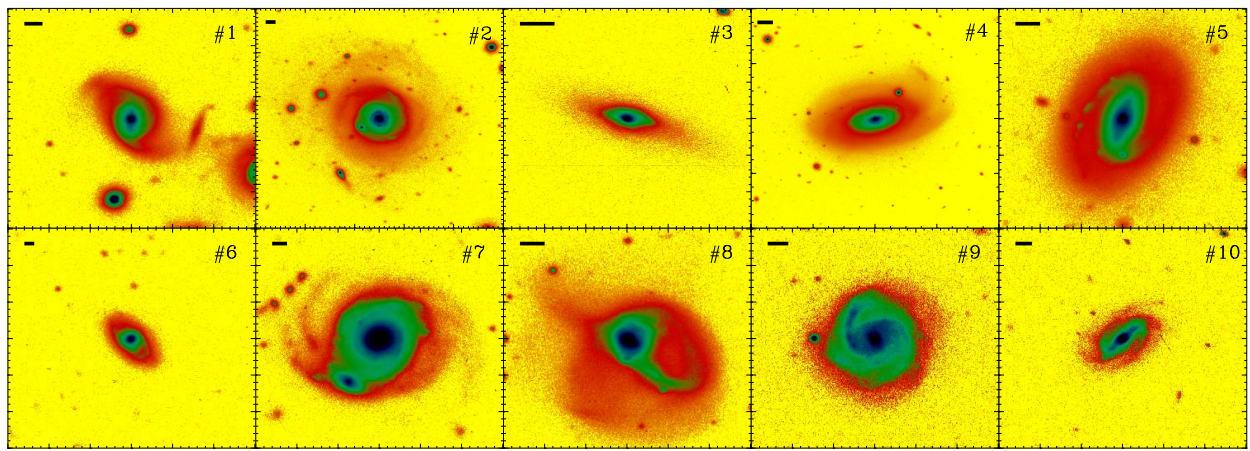}}
\caption{$R$-band NOT images of  the sample galaxies studied in  this work. The
  pseudo-color images  are displayed in  logarithmic scale to  show up
  low surface brightness-features. The scale  black bar on top left of
  each image  indicates a  physical angular separation  of 5  kpc. The
  number on the  top right for every image indicates  the galaxy ID as
  in Table~\ref{tab:tabla1}.  }
\label{fig:mosaico}
\end{figure*}

The  photometric  observations  were  carried  out  using  the  ALFOSC
instrument mounted at the 2.56-m  NOT telescope at La Palma. We obtain
deep and  high quality  $R$-band images using  the 2048  $\times$ 2048
back-illuminated CCD  with a  plate scale of  0.19 arcsec~pixel$^{-1}$
and  6.5 $\times$ 6.5  arcmin$^{2}$ field  of view.   The observations
were carried out during three runs (October 2006, April and May 2007).
In Table~\ref{tab:tabla1} we present  the list of observed objects and
other observational  details.  The exposure times were  chosen to keep
counts in  the detector linearity  regime, and to avoid  saturation in
the  bright central  parts of  the galaxies  to resolve  the different
structures present  in the galaxies.   Data reduction was made  in the
standard way  using IRAF\footnote{IRAF is distributed  by the National
  Optical  Astronomy  Observatories, operated  by  the Association  of
  Universities  for  Research in  Astronomy,  Inc., under  cooperative
  agreement  with   the  National  Science   Foundation.}.   The  bias
subtraction was done using a  master bias obtained at the beginning of
the night.   Flat-field corrections were applied using  sky flats.  We
have used our own script to automatize the alignment of the images and
removing cosmic rays at the same time.  Night conditions were clear in
general, and  2 nights (April 21st  and May 11th,  2007) were strictly
photometric.   Figure~\ref{fig:mosaico}   shows  a  mosaic   with  the
$R$-band images for our sample objects.

\section{Galaxy Photometric Decompositions}
\label{photdec}

The  structural parameters  of  the sample  galaxies  were derived  by
applying  a two-dimensional  photometric decomposition  to  the galaxy
images.    To   this   aim,   the  GASP2D   algorithm   developed   by
\citet{2008A&A...478..353M} was  used.  For  the sake of  clarity, and
because some  specific modifications  have been done  to the  code, we
will briefly  describe here the  main characteristics of  GASP2D.  The
galaxy surface-brightness distribution (SBD) was assumed to be the sum
of  different components  depending on  the morphological  features of
each   galaxy.    The   structural   components  considered   in   the
decomposition were the following:

The S\'ersic law \citep{1963BAAA....6...41S,1968adga.book.....S}, also
known  as the  $r^{1/n}$ law  or generalized  de Vaucouleurs  law, was
adopted to describe the surface brightness of the bulge component

\begin{equation} 
I_{\rm bulge}(r_{\rm bulge})=I_{\rm e}10^{-b_n\left[\left(\frac{r_{\rm bulge}}{r_{\rm e}} 
\right)^{\frac{1}{n}}-1\right]}, 
\label{eqn:bulge_surfbright} 
\end{equation} 
%
where  $r_{\rm  bulge}$  is  the  radius  measured  in  the  Cartesian
coordinates describing the reference system  of the bulge in the plane
of the  sky. $r_{\rm e}$, $I_{\rm  e}$, and $n$ are  the effective (or
half-light) radius, the surface brightness at $r_{\rm e}$, and a shape
parameter describing the curvature  of the SBD, respectively, and $b_n
= 2\,n-0.33$ \citep{1993MNRAS.265.1013C}.

The bulge isophotes are ellipses centered on the galaxy center ($x_0$,
$y_0$),  with constant  position angle  PA$_{\rm bulge}$  and constant
axial ratio $q_{\rm bulge}$.

The SBD of the disk component was assumed to follow an exponential law
\citep{1970ApJ...160..811F}

\begin{equation} 
I_{\rm disk}(r_{\rm disk}) = I_0\,e^{-\left(\frac{r_{\rm disk}}{h}\right)}, 
\label{eqn:disc_surfbright} 
\end{equation} 
%
where   $I_0$  and   $h$  are   the  central   surface brightness  and
scale-length  of  the  disk,  respectively.  The  disk  isophotes  are
ellipses  centered on  ($x_0$,  $y_0$), with  constant position  angle
PA$_{\rm disk}$ and constant axial ratio $q_{\rm disk}$.

The projected surface density of a three-dimensional Ferrers ellipsoid
(\citealt{Ferrers1877},  see  also  \citealt{aguerri09}) was  used  to
describe the SBD of bars

\begin{equation}
I_{\rm bar}(r_{\rm bar})=I_{\rm 0,bar}\left[1-\left(\frac{r_{\rm bar}}{a_{\rm bar}}\right)^2\right]^{n_{\rm bar}+0.5}; 
 \  r_{\rm bar} \le a_{\rm bar},
\end{equation}

where $I_{\rm  0,bar}$, $a_{\rm bar}$ and $n_{\rm  bar}$ represent the
central  surface brightness, length  and shape  parameter of  the bar,
respectively. Due  to the high  degree of degeneracy that  the $n_{\rm
  bar}$ parameter introduces during the  fit, we decided to keep it as
a fixed parameter  during the fitting process. The  default value used
was $n_{\rm bar}=2$  \citep[see][]{2005MNRAS.362.1319L}.  All the
bar  models  were  built  up   in  a  frame  of  generalized  ellipses
\citep{1990MNRAS.245..130A}.   Thus,  the   bar  reference  system  is
defined as

\begin{eqnarray} 
r_{\rm bar}&=&\left[(-(x-x_0) \sin{{\rm PA_{bar}}} + (y-y_0)
\cos{{\rm PA_{bar}}})^c  \right. \nonumber \\ & & 
- \left. { ((x-x_0) \cos{{\rm PA_{bar}}} + (y-y_0) \sin{{\rm PA_{bar}}})^c}\over{q_{\rm bar}^{\,c}}\right]^{1/c},
\label{eqn:bar_radius} 
\end{eqnarray} 
where $q_{\rm bar}$ and PA$_{\rm bar}$ are the axis ratio and position
angle of the  bar, respectively. The parameter $c$  controls the shape
of the isophotes.  A value  of $c=2$ corresponds to a perfect ellipse,
$c>2$ to a boxy shape and $c<2$ to a disky shape.

Since the central SBD of our objects is affected by the presence of an
AGN, we have modeled its contribution, $I_{\rm NC}(r_{\rm NC})$, by
means of a intensity scaled point spread function (PSF).

To derive  the photometric parameters  of the different  components we
fitted iteratively a model of the surface brightness
\begin{eqnarray}
I_{\rm model}(r) = I_{\rm bulge}(r_{\rm bulge}) \,+\,I_{\rm disk}(r_{\rm disk}) \,+\, I_{\rm bar}(r_{\rm bar}) \, \nonumber \\
+\, I_{\rm NC}(r_{\rm NC}),
\end{eqnarray}
to the  pixels of the  galaxy image, using a  non-linear least-squares
minimization based on a  robust Levenberg-Marquardt method. The actual
computation has been done using the MPFIT\footnote{The updated version
  of         this         code         is         available         on
  \url{http://cow.physics.wisc.edu/craigm/idl/idl.html}}      algorithm
implemented by C.~B.  Markwardt under the IDL
\footnote{Interactive  Data Language}  environment.  Each  image pixel
has  been weighted  according to  the variance  of its  total observed
photon counts due to the contribution  of both the galaxy and sky, and
determined assuming  photon noise  limitation and taking  into account
for the  detector readout  noise. The seeing  effects were  taken into
account by convolving the model  image with a circular Moffat PSF with
the FWHM measured from  stars in the galaxy image.

\begin{figure*}
{\includegraphics[angle=90,scale=0.5]{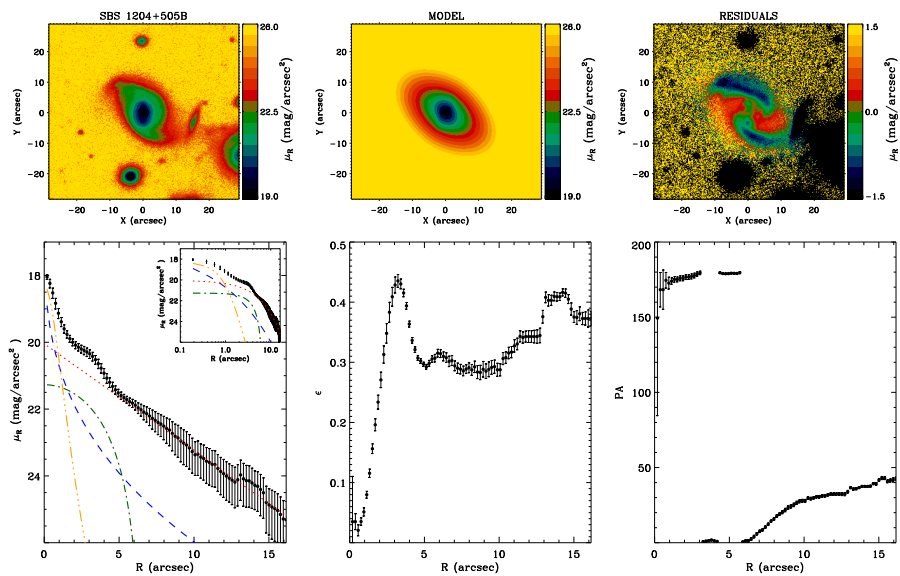}}
\caption{  {\it  Top  left:}  NOT  image  of  SDSS~J120655.63+501737.1
(SBS~1204+505B)  in the  $R$-band.  {\it  Top middle:}  galaxy model
 derived  from  the  GASP2D  fit considering  a  central  point-like
component  (fitted  with  the  PSF),  a  bulge, a  bar  and  a  disk
component.   {\it  Top  right:}  residuals image  derived  from  the
 subtraction of  the galaxy  model from the  NOT image.   {\it Bottom
 left:}      surface-brightness      distribution      of      SDSS
 J120655.63+501737.1.  Lines indicate the  contribution to the fit of
 the  different components  derived with  GASP2D: three-dotted-dashed
 line  for  the  central   component,  dashed  line  for  the  bulge,
 dotted-dashed line for the bar, and dotted line for the disk.  Upper
 inset shows the fit with a logarithmic scale for the galactocentric
distance.
{\it  Bottom  middle:} ellipticity  radial
 profile measured on the  galaxy image.  {\it Bottom right:} position
 angle (PA) radial profile measured on the galaxy image.  }
\label{fig:obj1}
\end{figure*}

\begin{figure*}
{\includegraphics[angle=90,scale=0.5]{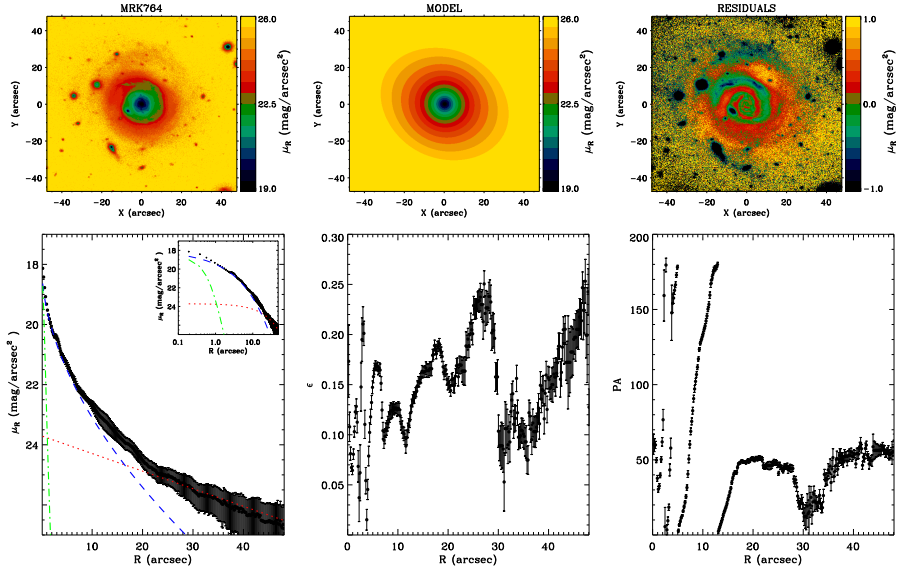}}
\caption{ J121600.04+124114.3 (\#2; Mrk~764). Panel distribution as in Figure~\ref{fig:obj1}.
}
\label{fig:obj2}
\end{figure*}

\begin{figure*}
{\includegraphics[angle=90,scale=0.5]{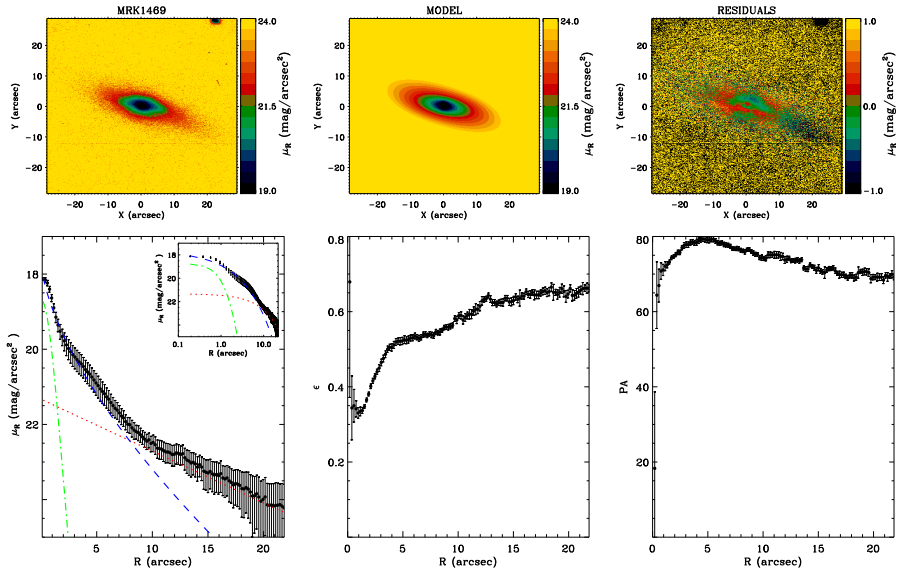}}
\caption{ J121607.08+504930.0 (\#3; Mrk~1469). Panel distribution as in Figure~\ref{fig:obj1}.
}
\label{fig:obj3}
\end{figure*}

\begin{figure*}
{\includegraphics[angle=90,scale=0.5]{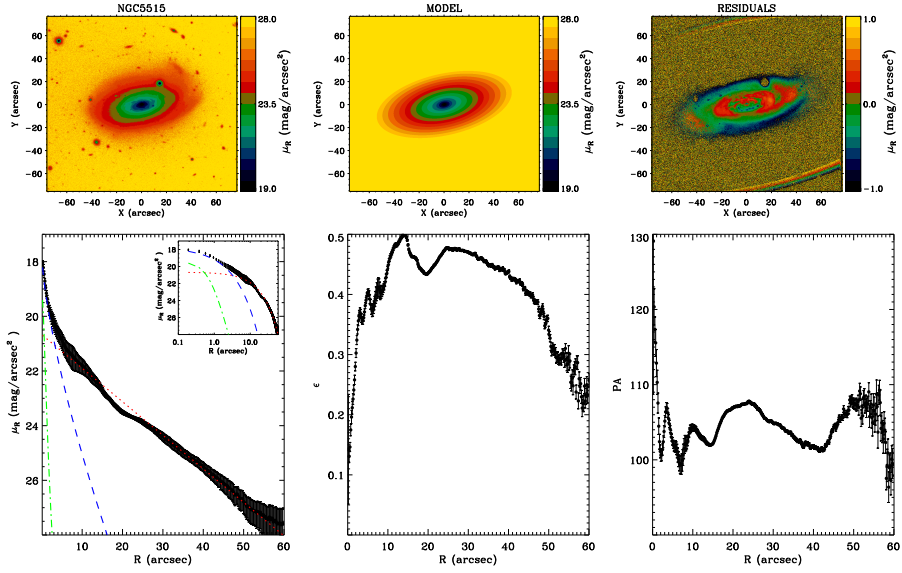}}
\caption{ J141238.14+391836.5 (\#4; NGC~5515). Panel distribution as in Figure~\ref{fig:obj1}.
}
\label{fig:obj4}
\end{figure*}

\begin{figure*}
{\includegraphics[angle=90,scale=0.5]{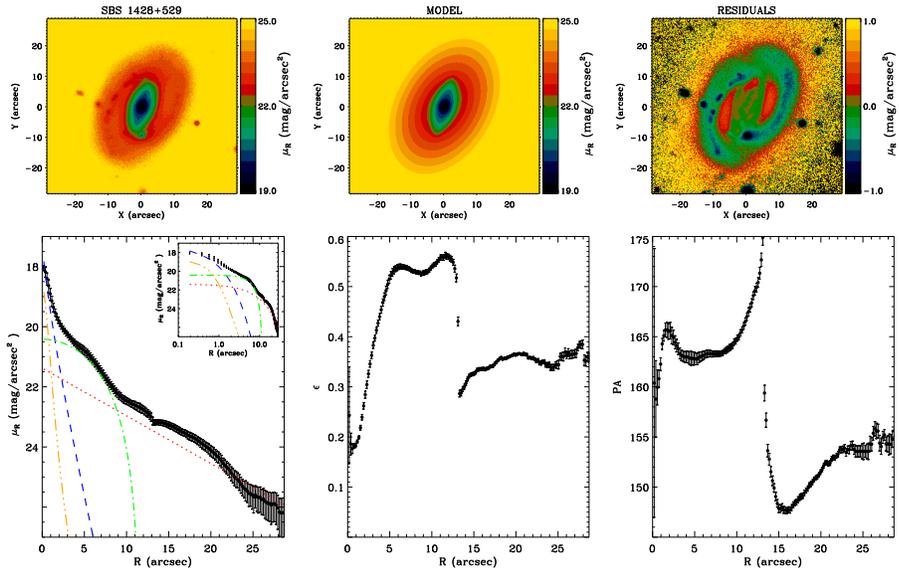}}
\caption{ J143031.18+524225.8 (\#5; SBS~1428+529). Panel distribution as in Figure~\ref{fig:obj1}.
}
\label{fig:obj5}
\end{figure*}

\begin{figure*}
{\includegraphics[angle=90,scale=0.5]{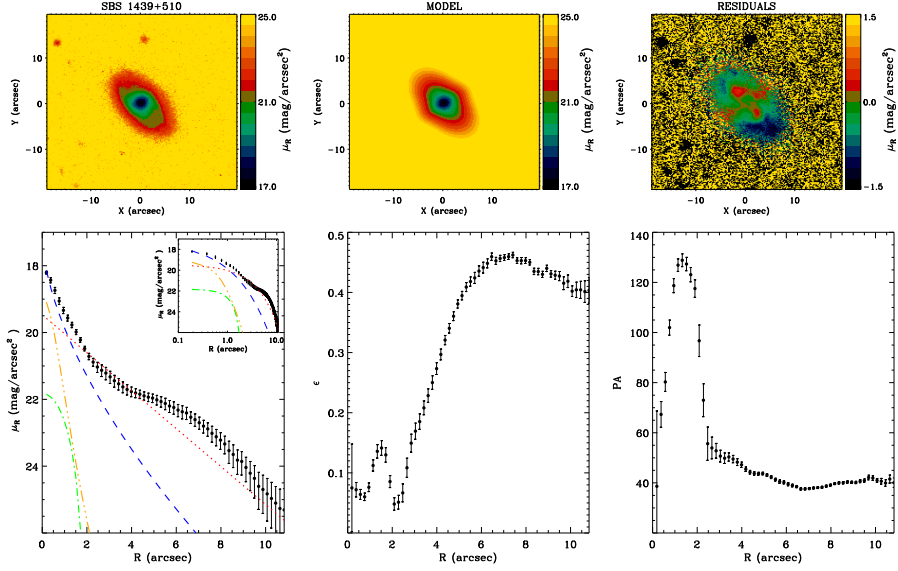}}
\caption{ J144049.35+505009.2 (\#6; SBS~1439+510). Panel distribution as in Figure~\ref{fig:obj1}.
}
\label{fig:obj6}
\end{figure*}

\begin{figure*}
{\includegraphics[angle=90,scale=0.5]{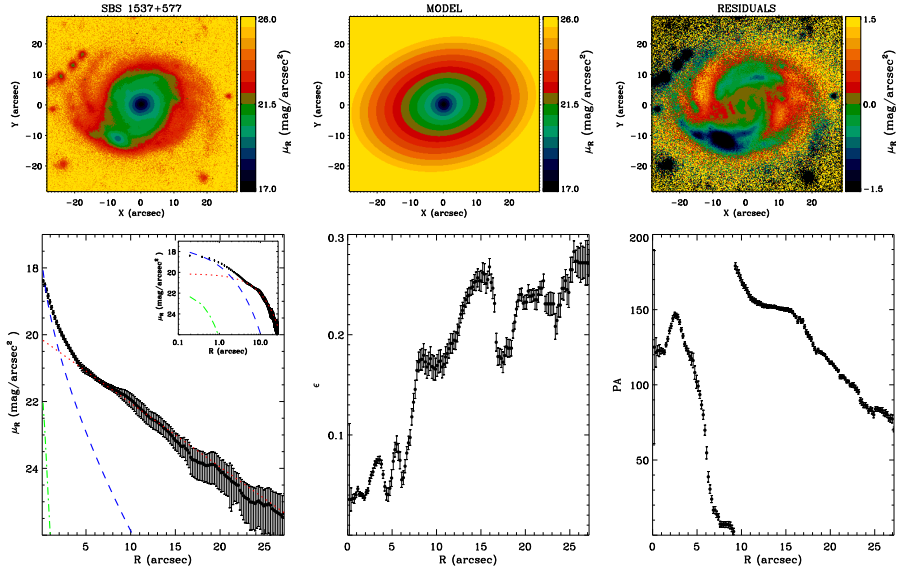}}
\caption{ J153810.05+573613.1 (\#7; SBS~1537+577). Panel distribution as in Figure~\ref{fig:obj1}.
}
\label{fig:obj7}
\end{figure*}

\begin{figure*}
{\includegraphics[angle=90,scale=0.5]{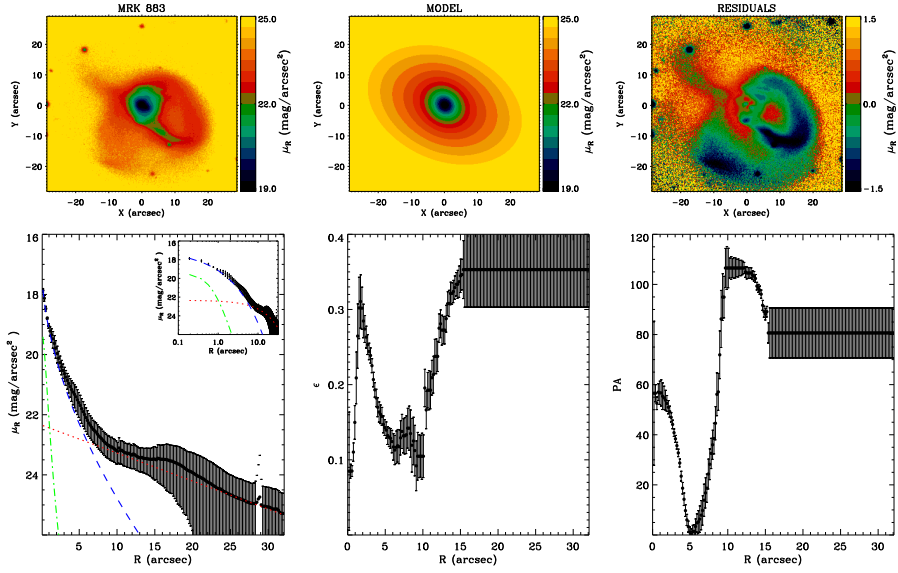}}
\caption{ J162952.88+242638.3 (\#8; Mrk~883).  Panel distribution as in Figure~\ref{fig:obj1}. 
}
\label{fig:obj8}
\end{figure*}

\begin{figure*}
{\includegraphics[angle=90,scale=0.5]{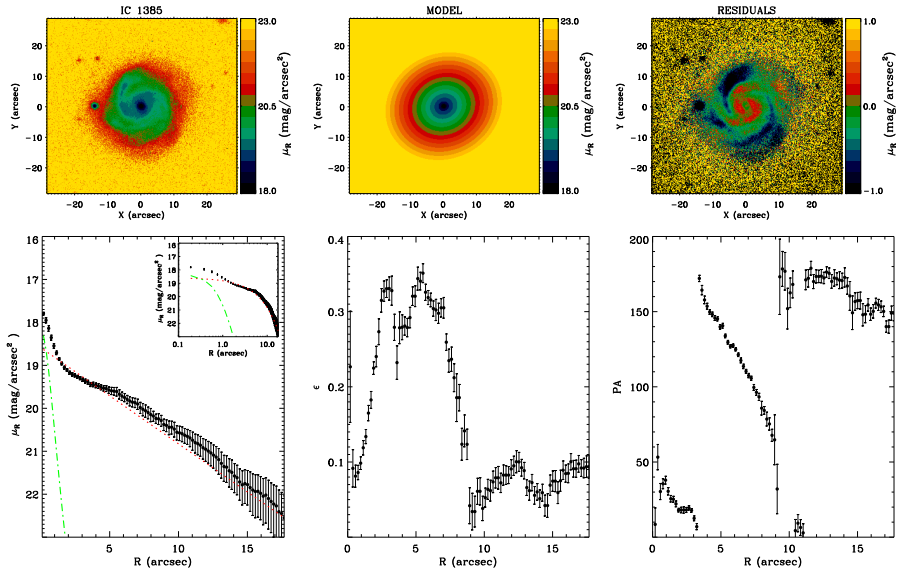}}
\caption{ J212851.19-010412.4 (\#9; IC~1385).   Panel distribution as in Figure~\ref{fig:obj1}.
}
\label{fig:obj9}
\end{figure*}

\begin{figure*}
{\includegraphics[angle=90,scale=0.5]{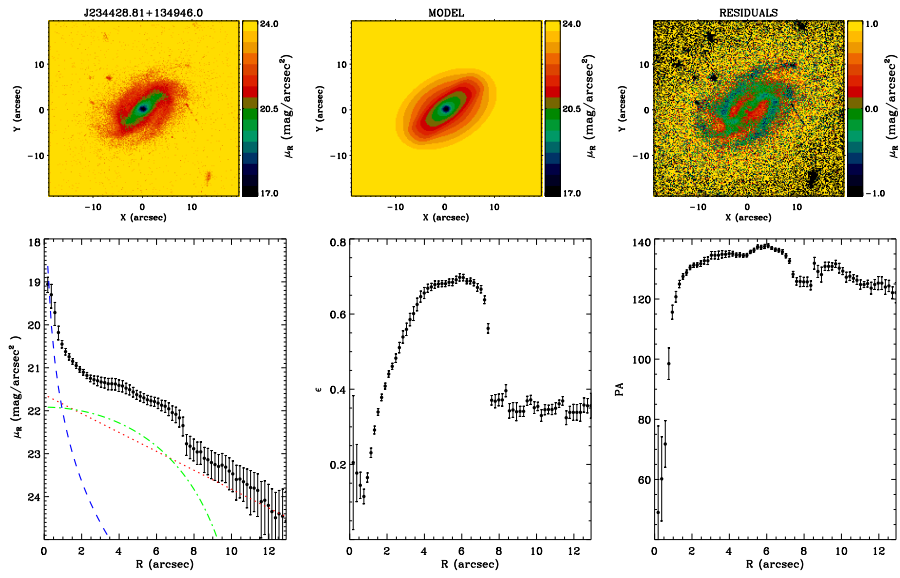}}
\caption{ J234428.81+134946.0 (\#10).  Panel distribution as in Figure~\ref{fig:obj1}.
}
\label{fig:obj10}
\end{figure*}

For each galaxy,  a model was fitted to the  SBD considering a central
point-like  component, a  bulge, a  bar  and a  disk component.     
Figures~\ref{fig:obj1} to ~\ref{fig:obj10}  show the GASP2D fits for
  each  galaxy  in  the  sample.   The  parameters  derived  for  the
structural  components, together  with  the $\chi^{2}$  values of  the
fits, are collected in Table~\ref{tab:tabla2}.

 The  formal  errors  obtained  from  the  $\chi^2$  minimization
  procedure are usually  not representative of the real  errors in the
  structural  parameters \citep{2008A&A...478..353M}.   Therefore, the
  errors  given  in Table  \ref{tab:tabla2}  were  obtained through  a
  series of Monte Carlo simulations.   A set of 500 images of galaxies
  with a  S\'ersic bulge, an exponential  disk, and a  central PSF was
  generated.  An  analogue set including the bar  component was also
  created.  The structural parameters  of the artificial galaxies were
  randomly chosen  among the ranges  obtained for our  sample galaxies
  (see Table~\ref{tab:tabla2}).   The simulated galaxies  were assumed
  to be at a distance of 261 Mpc, which corresponds to the mean of our
  galaxy   sample.    The  adopted   pixel   scale,   CCD  gain,   and
  read-out-noise were  chosen to mimic  the instrumental setup  of the
  photometric  observations.  Finally, a  background level  and photon
  noise were added to the artificial images to yield a signal-to-noise
  ratio  similar  to  that  of  the  observed  ones.   The  images  of
  artificial  galaxies  were analysed  with  GASP2D  as  if they  were
  real.  The  errors  on  the  fitted  parameters  were  estimated  by
  comparing the input and  measured values assuming they were normally
  distributed.  The mean and standard deviation of the relative errors
  of  the  artificial galaxies  were  adopted  as  the systematic  and
  typical errors for the observed galaxies.

 Another source of uncertainty  in this study is the PSF mismatch.
  This issue becomes crucial in this study since our modelling requires
  the assumption of a given PSF to perform the image convolution and to
  describe the AGN  component.  To get rid of  the PSF mismatch errors
  we have performed two different  tests.  First, we have repeated ten
  times  the fit  for every  galaxy allowing  the FWHM  of  the moffat
  function to vary within  5\%.  Second, we have also used a gaussian
  function with 5\% variation of the FWHM.  From the observed stars
  in the field,  we found that the gaussian function  is clearly a bad
  representation of  the PSF, however  we consider these errors  as an
  upper limit of the variation in the structural parameters due to the
  PSF  mismatch.   The final  errors  shown  in Table  \ref{tab:tabla2}
  represent the  combination in quadrature  of the Monte  Carlo errors
  derived previously and the errors due to PSF mismatch.

The reader is referred  to the individual galaxy description presented
in Sect.~\ref{results}, where details  on the morphology obtained from
NED\footnote{The NASA/IPAC Extragalactic Database (NED) is operated by
  the Jet  Propulsion Laboratory, California  Institute of Technology,
  under   contract   with   the   National   Aeronautics   and   Space
  Administration.}   and/or  derived in  this  work,  together with  a
discussion  on  the possible  host  bulge characterization  (classical
bulge or pseudobulge) are provided.

\section{Results}\label{results}

\subsection{Classical versus pseudobulges}
The  two-dimensional  photometric  decompositions  performed  on  deep
$R$-band images  allowed us  to determine the  nature of the  bulge of
each galaxy. We separated classical bulges from pseudobulges using the
prescriptions given by  \citet{2004ARA&A..42..603K}. The more  these
characteristics  apply, the safer the  classification  of pseudobulge
becomes. Among these characteristics we use the analysis of the SBD of
each  galaxy, the  value of  the S\'ersic  index (usually  bulges with
$n<2$ are considered pseudobulges; \citealt{2008AJ....136..773F}), the
ellipticities  of the  bulge and  disk, the  central  stellar velocity
dispersion  ($\sigma_{\star}$), and  by the  presence, or  absence, of
central structures from the  inspection of the $R$-band residuals 
derived using the best  fit.  Four galaxies  in our sample,  \#3, \#4, \#5  and \#7,
seem to  harbor a  pseudobulge, while 
for galaxies  \#1 {\bf, \#2}  and  \#6, our analysis  suggests a 
combination  of a classical bulge  and a  pseudobulge.  
For galaxies  \#9 and  \#10, the photometric decomposition did not allow 
us to determine their S\'ersic index, so we decided not to classify the 
nature of the bulge from this decomposition. Finally, galaxy \#8 seems  
to be a merger, and with the available data  we could  not derive a  
value for the  S\'ersic index, therefore we cannot  establish its bulge 
nature.  In summary, we found that {\bf  70\%$^{+10.1\%}_{-16.9\%}$} of 
the objects harbor pseudobulges or a combination of  classical 
bulge/pseudobulge,  which  suggests that pseudobulges might  be frequent  
in intermediate-type  AGN.   
{\bf In Paper I, double-peaked narrow emission lines were detected in 
the following intermediate-type AGN: \#1, \#3, \#4, \#5, and \#8. 
In this work, we could establish the nature of the bulge in four of them,
with the exception of object \#8, and found that they harbor either a pseudobulge 
 (\#3, \#4  and \#5) or a mix of both bulge-types (CB/PB for object \#1).}
Our morphological classification also shows that  most objects are  
early-type spirals, and our deep images allowed us to detect the presence 
of a bar in four of them. In the following section, we will explain in detail 
the  main properties of the individual galaxies.

\subsection{Notes on individual galaxies}

\subsubsection{{J120655.63+501737.1 (\#1)}}

The SBD  fit of  this galaxy is  presented in  Fig.~\ref{fig:obj1}. It
shows the presence of a 5.4 kpc bar.  The host galaxy probably presents
a combination of a  classical bulge/pseudobulge, since the photometric
decomposition provides us with a S\'ersic index $n$=2.8$\pm$1.2.  This
$n$ corresponds to  a classical bulge \citep{2004ARA&A..42..603K}, but
considering the errors it could  also be a pseudobulge. We cannot rule
out  a  pseudobulge  in  this  galaxy since  the  velocity  dispersion
($\sigma_{\star}$$\sim$ 90 km~s$^{-1}$) is  small and the galaxy has a
large  bar.    Both  properties  suggest   the  presence  of   a  more
rotationally   supported   component,    usually   associated   to   a
pseudobulge. We note  that this object has the  largest S\'ersic index
value in  the sample. The morphological  type of this  galaxy is T=2.9
(close to Sb) in the HyperLeda database
\footnote{\url{http://HyperLeda.univ-lyon1.fr}}. 
Since we found a bar component, we classify it as SBb. 

\subsubsection{{J121600.04+124114.3 (\#2)}}

The SBD  decomposition (see Fig.~\ref{fig:obj2})  suggests a classical
bulge  for this  galaxy, which  is supported by the absence of  a bar,  
and the different  ellipticities of the  bulge and disk components.  
The S\'ersic  index  $n=$1.8$\pm$0.6  favors  a pseudobulge ({\bf though} 
its  scatter is large) {\bf and the central velocity dispersion value 
is rather small ($\sigma_{\star}$$\sim$ 110 km~s$^{-1}$)}. This almost  
face-on galaxy is classified  as  Sb(r) in  the  NED  database. Its 
spiral  arms  are particularly clumpy in the inner parts (within the  
central 5 \arcsec\,$\sim$\,4 kpc)  and become very  diffuse towards 
the edge of the galaxy, the northern outer arm stretches out to the east.

\subsubsection{{J121607.08+504930.0 (\#3)}}

This  galaxy   probably  hosts  a  pseudobulge.   The  obtained  bulge
S\'ersic index  is $n$=1.6$\pm$0.5,  and the  similarity  between the
bulge  and disk ellipticities  ($\epsilon \sim  0.66$) along  with the
presence of a strong dust lane (see Fig.~\ref{fig:obj3}), indicates the
presence  of  a  rotating  component  in the  central  region  of  the
galaxy.  This  galaxy is  classified  as a  probable  SBb  in the  NED
database.  However, our  SBD  fit shows  no  indication of  a bar.  Thus,
we have re-classified this galaxy as an Sb.

\subsubsection{{J141238.14+391836.5 (\#4)}}

The SBD  analysis yields a  bulge S\'ersic index  $n$=1.4$\pm$0.8. The
bulge is  flat with an ellipticity  similar to that of  the disk, both
indicative  of  a  pseudobulge.   The  presence of  a  pseudobulge  is
strengthened by the central spiral arms shown in the residual image of
the  SBD decomposition  (see Fig.  \ref{fig:obj4}).  However,  we note
that  the central velocity  dispersion ({\bf $\sigma_{\star}$$\sim$ 195
km~s$^{-1}$})  is somewhat  large for  a pseudobulge.   This  galaxy is
classified as  Sab in both NED  and HyperLeda databases.
We found no indication of the presence of a bar in this galaxy.
  
\subsubsection{{J143031.18+524225.8 (\#5)}}

The SBD fit  indicates that this galaxy hosts  probably a pseudobulge.
The bulge S\'ersic index $n$=1.3$\pm$0.6, together with the similarity
in  bulge  and disk  ellipticities  ($\epsilon\,\sim$\,0.36), and  the
presence of a large bar ($a_{bar}=$7.0 kpc and $h=$4.7 kpc), reinforce
the idea  of secular processes acting  in this galaxy.  The galaxy has
two  spiral  arms that  are  barely  seen  in broad-band  images  (see
Fig.~\ref{fig:obj5}).  These  arms are  well  outlined  and clumpy  in
H$\alpha$   images  by   \citet{1995AJ....109..981W},   especially  the
northern one. Considering the presence of the bar and the shape of the
arms  traced by  H$\alpha$,  this  galaxy could  be  classified as  an
SBa-SBab galaxy.

\subsubsection{{J144049.35+505009.2 (\#6)}}

For this galaxy, the SBD  decomposition was done including a small bar
($a_{bar}$=2.8  kpc)  but   large  uncertainties  were  obtained.  Our
analysis shows that  this galaxy harbors both a  classical bulge and a
pseudobulge.   The S\'ersic  index obtained  is  $n$=1.6$\pm$0.8 but
compatible within  the errors with values $>$2.  The bulge ellipticity
indicates that  it is flat. However,  the presence of  a central small
bar might be  contaminating this result, as can be  seen in the galaxy
ellipticity profile (see  Fig.~\ref{fig:obj6}). Two very faint spiral 
arms  are seen in  the NOT  $R$-band image  over a  faint disk  and 
an elongation is  seen to the south  of the bulge. The  inspection of 
the $R$-band  image indicates  this could  be an  early-type  spiral.  
So, the morphological type is SBa.

\subsubsection{{J153810.05+573613.1 (\#7)}}

The SBD decomposition shows that this galaxy hosts a pseudobulge, with
a S\'ersic  index $n$=1.6$\pm$0.3. The central  velocity dispersion is
rather     low     ({\bf $\sigma_{\star}$$\sim$\, 91 \,km~s$^{-1}$}). From
Fig.~\ref{fig:obj7} we note that there are hints favoring the presence
of  central substructures.   Therefore, this  galaxy probably  hosts a
pseudobulge.  It is classified as Sc in NED since it displays open and
flocculent spiral arms.

\subsubsection{{J162952.88+242638.3 (\#8)}}

The SBD  fit obtained for  this galaxy must  be taken with  care since
this object is probably  an on-going merger (see Fig.~\ref{fig:obj8}).
Since the  morphology is disturbed and  the light profile  is not well
fitted with an elliptically averaged $r^{1/n}$ model, we classify this
galaxy as Peculiar.  For this reason, the  $M_{\rm BH}$ mass was
estimated considering both possibilities: classical bulge and a pseudobulge.

\subsubsection{{J212851.19-010412.4 (\#9)}}

The  effective radius  of the  bulge obtained  for this  galaxy always
converges   to   a   minimum   value   equal  to   the   PSF.    Thus,
Fig.~\ref{fig:obj9} shows  the best fit achieved  considering a bright
central component  and a bulgeless  galaxy.  Since the  S\'ersic index
cannot be determined,  it is uncertain whether it  harbors a classical
bulge or  a pseudobulge. We found no  morphological classification for
this galaxy in NED, nor  in the HyperLeda database.  Inspection of the
NOT  $R$-band image  shows  that it  is  a spiral  galaxy seen  almost
face-on. 
From  this image, it is not  clear if the bulge  is large and boxy,  nor  if  it  harbors  a  bar.
The  spiral  arms  are  a  bit flocculent. The eastern arm of the galaxy is brighter than its central
parts-, then it becomes very  diffuse and seems to bifurcate.  For the
western side of the galaxy, two  very similar arms seem to emerge from
the  disk.  Considering their  degree of  tightness, we  classify this
galaxy  as an  Sab.

\subsubsection{{J234428.81+134946.0 (\#10)}}

For this  galaxy, the SBD decomposition  includes a 9.9 kpc  bar and a
disk with $h$\,=\,5.1\,kpc, but the bulge cannot be resolved since its
effective radius always converges to a minimum value equal to the PSF.
For  this  reason,  the  value   for  the  S\'ersic  index  cannot  be
determined, nor  whether it  has a classical  bulge or  a pseudobulge.
This  galaxy has  a  de Vaucouleurs'  type  of 4.1  $\pm$  5.0 in  the
HyperLEDA database.  This corresponds  to a morphological type of Sbc,
though the error does not allow a precise classification. According to
our NOT  image and  the SBD decomposition  (see Fig.~\ref{fig:obj10}),
this is a barred galaxy.  Arms are very faint, tight and flocculent. A
very  faint  arm  is  seen  in  the northern  part  extending  to  the
north-west. We classify this galaxy as a SBab.

\subsection{$M_{BH}$ estimates}

Using the  $\sigma_{\star}$ calculated  in Paper I,  we obtained {\bf the BH mass} using  either  the  $M_{\rm  BH}$-$\sigma_{\star}$ relation  given by  \citet{2005SSRv..116..523F}  for classical  bulges ($\alpha$\, = \,8.06$\pm$0.67, \,$\beta$\, = \,4.86$\pm$0.43) or {\bf the one given by} \citet{2008MNRAS.386.2242H}, ($\alpha$\, = \,7.50$\pm$0.18, \,$\beta$\, = \,4.5$\pm$1.3) for pseudobulges. In the cases  where the bulges are probably a combination of  both  classical  and  pseudobulge (CB/PB) we  estimated the $M_{\rm BH}^{\sigma_\star}$ using both correlations.  The results of these $M_{\rm BH}^{\sigma_\star}$ estimations are given in Table~\ref{tab:tabla3}, see Col(4) and (5). Also in this table the {\bf BH} mass estimates ($M_{\rm BH}$) obtained in Paper  I are presented in Col (6). The $M_{\rm BH}^{\sigma_{\star}}$ has a range of {\bf 5.69$\pm$0.21\,$<$\,log\,$M_{\rm BH}^{\sigma_{\star}}$\,$<$\,8.09$\pm$0.24.} 

Figure \ref{fig:masscomp} shows the  comparison between the  $M_{\rm BH}^{\sigma_{\ast}}$ masses obtained depending on the nature of the bulge with the  $M_{\rm BH}$ {\bf estimated in Paper I. In this figure, we have marked in the upper and lower panels a region defined by dotted lines that corresponds to 3 and 1/3 times the $M_{\rm BH}=M_{\rm BH}^{\sigma_{\star}}$ relation. In the upper panel we show our results for seven out of ten galaxies that we have found harbor a PB or a mix of both, i.e. CB/PB. We also show the estimations obtained for the other three galaxies that still have unknown bulge type assuming that they could harbor a PB.  From this panel, we see that object \#8 lies well within the dotted lines, object \#10 lies marginally but object \#9 clearly lies out.
Now, considering only the objects with bulge-type found in this work, we find that objects \#4, \#5 and \#6 fall inside the dotted region. The remaining four objects \#1,\#2,\#3 and \#7 are out of the dotted region.  Furthermore, in the lower panel we show the estimates obtained for objects that were found to harbor a mix of CB/PB, i.e. \#1, \#2 and \#6. In this case, $M_{\rm BH}^{\sigma_{\star}}$ was estimated using \citet{2005SSRv..116..523F}. It is interesting to note that two of them lie marginally inside the dotted region and one lies completely out of it (\#6). Regarding the three unknown bulge-type objects, we see that object \#10 lies inside the dotted region, object \#9 could be considered a marginal case and object \#8 lies out of the region.}

\section{Discussion and conclusions}
\label{discussion}

In this paper, we derived the R-band host-bulge structural parameters of a sample of 10 intermediate-type AGN. Through a detailed two-dimensional photometric decomposition analysis we find that most of the host galaxies are early-type spirals and four of them have a bar. This analysis, together {\bf with a careful morphological analysis} and the velocity dispersion obtained in Paper 1, {\bf allowed us to determine the nature of the bulge in seven out of ten intermediate-type AGN. In these seven objects, we find that all of them harbor a pseudobulge or a combination of both bulge-types (CB/PB). To our knowledge, this is the first time that such a detailed study on the nature of the bulge-type is done to a sample of intermediate-type AGN. Although the sample is not extensive, our work strongly suggests that pseudobulges are actually frequent in this class of objects.}

In  Paper  I, we  suggested  that  intermediate-type AGN  should  be analyzed separately from Type 1 or 2 AGN due to the high fraction of narrow double-peaked sources found in our sample (50\%$\pm$14.4\%). In this work, we have also find that narrow double-peaked  emission lines are more  frequently  found in galaxies harboring a pseudobulge  or  a combination of classical  bulge/pseudobulge  (four out of {\bf seven galaxies with bulge classification are double-peaked, 57\%$^{+14.9\%}_{-18.1\%}$ harbor a PB or CB/PB)}.

On the basis of the bulge nature, we calculated $M_{BH}^{\sigma_\star}$ using the empirical relation given by \citet{2008MNRAS.386.2242H} for pseudobulges, and also the relation given by \citet{2005SSRv..116..523F} for classical bulges {\bf when the bulge-type resulted to be a mix of both. We have also used the M$_{BH}$ estimations given in Paper 1 that were obtained with the scaling relations given by \citet{2006ApJ...641..689V}. 
The black hole mass range obtained with the three methods is {\bf 5.69$\pm$0.21 $<$ $\log M_{BH}$ $<$ 8.09$\pm$0.24}.
The three tested methods yield no systematically different results for the range of masses: the range for classical bulges is 6.26$\pm$0.04 $<$ $\log{M_{BH}^{\sigma_{\star}}}$ $<$ 8.09$\pm$0.24; for pseudobulges 5.69$\pm$0.21 $<$ $\log{M_{BH}^{\sigma_{\star}}}$ $<$ 7.38$\pm$0.32; and with the scaling relations yields 6.54$\pm$0.16 $<$ $\log{M_{BH}}<$7.81$\pm$0.14. It should be noted that, since we have showed that pseudobulges are frequent in intermediate type-AGN, and taken into account that the M$_{BH}-\sigma*$ relation for classical bulges gives systematically higher masses than the one for pseudobulges, in principle one should use the pseudobulge M$_{BH}-\sigma*$ relation in order to not overestimate the black hole mass. 

However, comparing our $M_{BH}^{\sigma_{\star}}$ estimates with the $M_{BH}$ obtained using the scaling relations, we find that only four out of the ten  (40$^{+16}_{-13}\%$) galaxies (\#5, \#6, \#8 and \#10) are compatible within 1-$\sigma$ errors and one more (\#4) is compatible within 3-$\sigma$ (that is, 50$\pm20\%$ in total). \citet{2012JPhCS.372a2008P} advised that black hole mass estimations using the scaling relations can be reliable within a factor of $\sim$3. When this criterium is applied, the masses obtained with $M_{BH}^{\sigma_{\star}}$ relations and $M_{BH}$ are compatible for eight (all except \#3 and \#7) of the galaxies in the sample (80$^{+7}_{-17}\%$) . If this criterium is valid for our sample, our results show that the scaling relations derived from Type 1 AGN \citep[e.g.,][]{2010ApJ...716..269W,2011ApJ...726...59B} are the same for intermediate-type AGN.

From our seven classified bulges we could specify that four of them (57$^{+15}_{-18}\%$) harbor certainly pseudobulges (\#3, \#4, \#5, \#7). For these, only half (50$\pm20\%$) of them (\#4 and \#5) present masses compatible within 3-$\sigma$ for both methods (upper panel figure~\ref{fig:masscomp}). The same result is obtained when the factor-3 criterium in the mass determination is applied. The remaining three galaxies (\#1, \#2 and \#6) are compatible with both bulge types being classical and pseudobulge. Assuming that these three objects do harbor a pseudobulge, only one out three (\#6) presents compatible masses considering both the 3-$\sigma$ and the factor-3 criteria. On the other hand, considering them as classical bulges, we find that none have compatible masses within 3-$\sigma$ errors, but two of them (\#1 and \#2) do actually have compatible masses if the factor-3 criterium is applied (lower panel figure~\ref{fig:masscomp}). 

In summary, when the type of the bulge is taken into account, only three (depending on both things: the criterium applied to the uncertainties and the type of bulge for the CB/PB galaxies) out of the seven (43$^{+18}_{-15}\%$) galaxies of the sample have compatible $M_{BH}^{\sigma*}$ and $M_{BH}$.
Nothing can be concluded for galaxies \#8, \#9 and \#10 since the nature of their bulge is unknown. We also find that the black hole mass estimates based on the $M_{BH}-\sigma*$ relation for pseudobulges is not compatible in 50$\pm20\%$ of the objects.Therefore, these objects would support the result found by \citet{2011Natur.469..374K} since their velocity dispersion does not correlate with the black hole mass. However, our results are based on a small number of objects, and therefore we stress the importance to perform similar detailed analysis in which the type of bulge can be determine using larger samples. }

\acknowledgements

We want to thank the anonymous referee for useful comments and 
suggestions. We also thank Dr. D. Clark who carefully read the final
version of the manuscript. EB  acknowledges financial support from 
UNAM-DGAPA-PAPIIT through grant  IN116211. JMA  is partially funded by the Spanish MICINN
under the Consolider-Ingenio 2010 Program grant CSD2006-00070 and also
by  the  grants  AYA2007-67965-C03-01  and  AYA2010-21887-C04-04.  IFC
thanks  the financial  support from  CONACYT grant  0133520 and IPN-SIP 
grant 20121700.  The data presented here were obtained with ALFOSC, which is provided by 
the Instituto de Astrof\'\i sica de Andaluc\'\i a (IAA) under a joint 
agreement with the University of Copenhagen and NOTSA.
Funding for the SDSS and SDSS-II has been provided by the Alfred P. Sloan Foundation, the Participating Institutions, the National Science Foundation, the U.S. Department of Energy, the National Aeronautics and Space Administration, the Japanese Monbukagakusho, the Max Planck Society, and the Higher Education Funding Council for England. The SDSS Web Site is http://www.sdss.org/. 
This research has  made use of  the NASA/IPAC Extragalactic  Database (NED)
which  is  operated  by  the  Jet  Propulsion  Laboratory,  California
Institute of Technology, under  contract with the National Aeronautics
and Space  Administration. We acknowledge  the usage of  the HyperLeda
database (http://leda.univ-lyon1.fr).

\bibliography{syintermedias-pII_nov2012}

\begin{thebibliography}{40}
\expandafter\ifx\csname natexlab\endcsname\relax\def\natexlab#1{#1}\fi

\bibitem[{{Aguerri} {et~al.}(2009){Aguerri}, {M{\'e}ndez-Abreu}, \&
  {Corsini}}]{aguerri09}
{Aguerri}, J.~A.~L., {M{\'e}ndez-Abreu}, J., \& {Corsini}, E.~M. 2009, \aap,
  495, 491

\bibitem[{{Andredakis} {et~al.}(1995){Andredakis}, {Peletier}, \&
  {Balcells}}]{1995MNRAS.275..874A}
{Andredakis}, Y.~C., {Peletier}, R.~F., \& {Balcells}, M. 1995, \mnras, 275,
  874

\bibitem[{{Andredakis} \& {Sanders}(1994)}]{1994MNRAS.267..283A}
{Andredakis}, Y.~C. \& {Sanders}, R.~H. 1994, \mnras, 267, 283

\bibitem[{{Athanassoula} {et~al.}(1990){Athanassoula}, {Morin}, {Wozniak},
  {Puy}, {Pierce}, {Lombard}, \& {Bosma}}]{1990MNRAS.245..130A}
{Athanassoula}, E., {Morin}, S., {Wozniak}, H., {Puy}, D., {Pierce}, M.~J.,
  {Lombard}, J., \& {Bosma}, A. 1990, \mnras, 245, 130

\bibitem[{{Beifiori} {et~al.}(2009){Beifiori}, {Sarzi}, {Corsini}, {Dalla
  Bont{\`a}}, {Pizzella}, {Coccato}, \& {Bertola}}]{2009ApJ...692..856B}
{Beifiori}, A., {Sarzi}, M., {Corsini}, E.~M., {Dalla Bont{\`a}}, E.,
  {Pizzella}, A., {Coccato}, L., \& {Bertola}, F. 2009, \apj, 692, 856

\bibitem[{{Bennert} {et~al.}(2011){Bennert}, {Auger}, {Treu}, {Woo}, \&
  {Malkan}}]{2011ApJ...726...59B}
{Bennert}, V.~N., {Auger}, M.~W., {Treu}, T., {Woo}, J.-H., \& {Malkan}, M.~A.
  2011, \apj, 726, 59

\bibitem[{{Caon} {et~al.}(1993){Caon}, {Capaccioli}, \&
  {D'Onofrio}}]{1993MNRAS.265.1013C}
{Caon}, N., {Capaccioli}, M., \& {D'Onofrio}, M. 1993, \mnras, 265, 1013

\bibitem[{{Ferrarese} \& {Ford}(2005)}]{2005SSRv..116..523F}
{Ferrarese}, L. \& {Ford}, H. 2005, \ssr, 116, 523

\bibitem[{{Ferrarese} \& {Merritt}(2000)}]{2000ApJ...539L...9F}
{Ferrarese}, L. \& {Merritt}, D. 2000, \apjl, 539, L9

\bibitem[{Ferrers(1877)}]{Ferrers1877}
Ferrers, N.~M. 1877, Quart. J. Pure Appl. Math., 14, 1

\bibitem[{{Fisher} \& {Drory}(2008)}]{2008AJ....136..773F}
{Fisher}, D.~B. \& {Drory}, N. 2008, \aj, 136, 773

\bibitem[{{Freeman}(1970)}]{1970ApJ...160..811F}
{Freeman}, K.~C. 1970, \apj, 160, 811

\bibitem[{{Ganda} {et~al.}(2007){Ganda}, {Peletier}, {McDermid},
  {Falc{\'o}n-Barroso}, {de Zeeuw}, {Bacon}, {Cappellari}, {Davies},
  {Emsellem}, {Krajnovi{\'c}}, {Kuntschner}, {Sarzi}, \& {van de
  Ven}}]{2007MNRAS.380..506G}
{Ganda}, K., {Peletier}, R.~F., {McDermid}, R.~M., {Falc{\'o}n-Barroso}, J.,
  {de Zeeuw}, P.~T., {Bacon}, R., {Cappellari}, M., {Davies}, R.~L.,
  {Emsellem}, E., {Krajnovi{\'c}}, D., {Kuntschner}, H., {Sarzi}, M., \& {van
  de Ven}, G. 2007, \mnras, 380, 506

\bibitem[{{Gebhardt} {et~al.}(2000){Gebhardt}, {Bender}, {Bower}, {Dressler},
  {Faber}, {Filippenko}, {Green}, {Grillmair}, {Ho}, {Kormendy}, {Lauer},
  {Magorrian}, {Pinkney}, {Richstone}, \& {Tremaine}}]{2000ApJ...539L..13G}
{Gebhardt}, K., {Bender}, R., {Bower}, G., {Dressler}, A., {Faber}, S.~M.,
  {Filippenko}, A.~V., {Green}, R., {Grillmair}, C., {Ho}, L.~C., {Kormendy},
  J., {Lauer}, T.~R., {Magorrian}, J., {Pinkney}, J., {Richstone}, D., \&
  {Tremaine}, S. 2000, \apjl, 539, L13

\bibitem[{{Graham}(2008)}]{2008ApJ...680..143G}
{Graham}, A.~W. 2008, \apj, 680, 143

\bibitem[{{Graham} \& {Li}(2009)}]{2009ApJ...698..812G}
{Graham}, A.~W. \& {Li}, I.-h. 2009, \apj, 698, 812

\bibitem[{{Greene} {et~al.}(2008){Greene}, {Ho}, \&
  {Barth}}]{2008ApJ...688..159G}
{Greene}, J.~E., {Ho}, L.~C., \& {Barth}, A.~J. 2008, \apj, 688, 159

\bibitem[{{G{\"u}ltekin} {et~al.}(2009){G{\"u}ltekin}, {Richstone}, {Gebhardt},
  {Lauer}, {Tremaine}, {Aller}, {Bender}, {Dressler}, {Faber}, {Filippenko},
  {Green}, {Ho}, {Kormendy}, {Magorrian}, {Pinkney}, \&
  {Siopis}}]{2009ApJ...698..198G}
{G{\"u}ltekin}, K., {Richstone}, D.~O., {Gebhardt}, K., {Lauer}, T.~R.,
  {Tremaine}, S., {Aller}, M.~C., {Bender}, R., {Dressler}, A., {Faber}, S.~M.,
  {Filippenko}, A.~V., {Green}, R., {Ho}, L.~C., {Kormendy}, J., {Magorrian},
  J., {Pinkney}, J., \& {Siopis}, C. 2009, \apj, 698, 198

\bibitem[{{Hu}(2008)}]{2008MNRAS.386.2242H}
{Hu}, J. 2008, \mnras, 386, 2242

\bibitem[{{Kauffmann} {et~al.}(1993){Kauffmann}, {White}, \&
  {Guiderdoni}}]{1993MNRAS.264..201K}
{Kauffmann}, G., {White}, S.~D.~M., \& {Guiderdoni}, B. 1993, \mnras, 264, 201

\bibitem[{{Kormendy} {et~al.}(2011){Kormendy}, {Bender}, \&
  {Cornell}}]{2011Natur.469..374K}
{Kormendy}, J., {Bender}, R., \& {Cornell}, M.~E. 2011, \nat, 469, 374

\bibitem[{{Kormendy} \& {Kennicutt}(2004)}]{2004ARA&A..42..603K}
{Kormendy}, J. \& {Kennicutt}, Jr., R.~C. 2004, \araa, 42, 603

\bibitem[{{Laurikainen} {et~al.}(2005){Laurikainen}, {Salo}, \&
  {Buta}}]{2005MNRAS.362.1319L}
{Laurikainen}, E., {Salo}, H., \& {Buta}, R. 2005, \mnras, 362, 1319

\bibitem[{{Laurikainen} {et~al.}(2007){Laurikainen}, {Salo}, {Buta}, \&
  {Knapen}}]{2007MNRAS.381..401L}
{Laurikainen}, E., {Salo}, H., {Buta}, R., \& {Knapen}, J.~H. 2007, \mnras,
  381, 401

\bibitem[{{Mehlert} {et~al.}(2003){Mehlert}, {Thomas}, {Saglia}, {Bender}, \&
  {Wegner}}]{2003A&A...407..423M}
{Mehlert}, D., {Thomas}, D., {Saglia}, R.~P., {Bender}, R., \& {Wegner}, G.
  2003, \aap, 407, 423

\bibitem[{{M{\'e}ndez-Abreu} {et~al.}(2008{\natexlab{a}}){M{\'e}ndez-Abreu},
  {Aguerri}, {Corsini}, \& {Simonneau}}]{2008A&A...478..353M}
{M{\'e}ndez-Abreu}, J., {Aguerri}, J.~A.~L., {Corsini}, E.~M., \& {Simonneau},
  E. 2008{\natexlab{a}}, \aap, 478, 353

\bibitem[{{M{\'e}ndez-Abreu} {et~al.}(2008{\natexlab{b}}){M{\'e}ndez-Abreu},
  {Corsini}, {Debattista}, {De Rijcke}, {Aguerri}, \&
  {Pizzella}}]{2008ApJ...679L..73M}
{M{\'e}ndez-Abreu}, J., {Corsini}, E.~M., {Debattista}, V.~P., {De Rijcke}, S.,
  {Aguerri}, J.~A.~L., \& {Pizzella}, A. 2008{\natexlab{b}}, \apjl, 679, L73

\bibitem[{{M{\'e}ndez-Abreu} {et~al.}(2010){M{\'e}ndez-Abreu}, {Simonneau},
  {Aguerri}, \& {Corsini}}]{2010A&A...521A..71M}
{M{\'e}ndez-Abreu}, J., {Simonneau}, E., {Aguerri}, J.~A.~L., \& {Corsini},
  E.~M. 2010, \aap, 521, A71+

\bibitem[{{Morelli} {et~al.}(2008){Morelli}, {Pompei}, {Pizzella},
  {M{\'e}ndez-Abreu}, {Corsini}, {Coccato}, {Saglia}, {Sarzi}, \&
  {Bertola}}]{2008MNRAS.389..341M}
{Morelli}, L., {Pompei}, E., {Pizzella}, A., {M{\'e}ndez-Abreu}, J., {Corsini},
  E.~M., {Coccato}, L., {Saglia}, R.~P., {Sarzi}, M., \& {Bertola}, F. 2008,
  \mnras, 389, 341

\bibitem[{{Peterson}(2012)}]{2012JPhCS.372a2008P}
{Peterson}, B.~M. 2012, Journal of Physics Conference Series, 372, 012008

\bibitem[{{S{\'e}rsic}(1963)}]{1963BAAA....6...41S}
{S{\'e}rsic}, J.~L. 1963, Boletin de la Asociacion Argentina de Astronomia La
  Plata Argentina, 6, 41

\bibitem[{{S{\'e}rsic}(1968)}]{1968adga.book.....S}
---. 1968, {Atlas de galaxias australes} ({Cordoba, Argentina: Observatorio
  Astronomico, 1968})

\bibitem[{{Stephens} {et~al.}(2003){Stephens}, {Frogel}, {DePoy}, {Freedman},
  {Gallart}, {Jablonka}, {Renzini}, {Rich}, \& {Davies}}]{2003AJ....125.2473S}
{Stephens}, A.~W., {Frogel}, J.~A., {DePoy}, D.~L., {Freedman}, W., {Gallart},
  C., {Jablonka}, P., {Renzini}, A., {Rich}, R.~M., \& {Davies}, R. 2003, \aj,
  125, 2473

\bibitem[{{Thomas} \& {Davies}(2006)}]{2006MNRAS.366..510T}
{Thomas}, D. \& {Davies}, R.~L. 2006, \mnras, 366, 510

\bibitem[{{Thomas} {et~al.}(2005){Thomas}, {Maraston}, {Bender}, \& {Mendes de
  Oliveira}}]{2005ApJ...621..673T}
{Thomas}, D., {Maraston}, C., {Bender}, R., \& {Mendes de Oliveira}, C. 2005,
  \apj, 621, 673

\bibitem[{{Vestergaard} \& {Peterson}(2006)}]{2006ApJ...641..689V}
{Vestergaard}, M. \& {Peterson}, B.~M. 2006, \apj, 641, 689

\bibitem[{{Weistrop} {et~al.}(1995){Weistrop}, {Hintzen}, {Liu}, {Lowenthal},
  {Cheng}, {Oliversen}, {Brown}, \& {Woodgate}}]{1995AJ....109..981W}
{Weistrop}, D., {Hintzen}, P., {Liu}, C., {Lowenthal}, J., {Cheng}, K.,
  {Oliversen}, R., {Brown}, L., \& {Woodgate}, B. 1995, \aj, 109, 981

\bibitem[{{Woo} {et~al.}(2010){Woo}, {Treu}, {Barth}, {Wright}, {Walsh},
  {Bentz}, {Martini}, {Bennert}, {Canalizo}, {Filippenko}, {Gates}, {Greene},
  {Li}, {Malkan}, {Stern}, \& {Minezaki}}]{2010ApJ...716..269W}
{Woo}, J.-H., {Treu}, T., {Barth}, A.~J., {Wright}, S.~A., {Walsh}, J.~L.,
  {Bentz}, M.~C., {Martini}, P., {Bennert}, V.~N., {Canalizo}, G.,
  {Filippenko}, A.~V., {Gates}, E., {Greene}, J., {Li}, W., {Malkan}, M.~A.,
  {Stern}, D., \& {Minezaki}, T. 2010, \apj, 716, 269

\bibitem[{{Wyse} {et~al.}(1997){Wyse}, {Gilmore}, \&
  {Franx}}]{1997ARA&A..35..637W}
{Wyse}, R.~F.~G., {Gilmore}, G., \& {Franx}, M. 1997, \araa, 35, 637

\bibitem[{{Zoccali} {et~al.}(2003){Zoccali}, {Renzini}, {Ortolani}, {Greggio},
  {Saviane}, {Cassisi}, {Rejkuba}, {Barbuy}, {Rich}, \&
  {Bica}}]{2003A&A...399..931Z}
{Zoccali}, M., {Renzini}, A., {Ortolani}, S., {Greggio}, L., {Saviane}, I.,
  {Cassisi}, S., {Rejkuba}, M., {Barbuy}, B., {Rich}, R.~M., \& {Bica}, E.
  2003, \aap, 399, 931

\end{thebibliography}

\begin{figure*}
\includegraphics[scale=0.6, angle=270]{Fig12a.ps}
\includegraphics[scale=0.6, angle=270]{Fig12b.ps}
\caption{{\it Upper panel:} $M_{\rm BH}^{\sigma_{\star}}$ derived for pseudobulges using \citet{2008MNRAS.386.2242H} compared to $M_{\rm  BH}$} estimates derived using the scaling relations given by \citet{2006ApJ...641..689V}. {\it Lower panel:} $M_{\rm BH}^{\sigma_{\star}}$ derived for classical bulges using \citet{2005SSRv..116..523F} compared to {\bf $M_{\rm  BH}$, see text.
}
\label{fig:masscomp}
\end{figure*}

\begin{deluxetable}{cccccccccc}
\tabletypesize{\scriptsize}
\tablecolumns{9}
\tablewidth{0pc}
\tablecaption{{Galaxy Sample and Observations}\label{tab:tabla1}}
\tablehead{
\colhead{Galaxy} & \colhead{SDSS} & \colhead{Other} &\colhead{NOT-ALFOSC} & \colhead{Exp-time} & \colhead{Seeing} \\
\colhead{\#}     & \colhead{ID}   & \colhead{name} &\colhead{Date-Obs}   & \colhead{(s)}      & \colhead{(\arcsec)} \\
\colhead{(1)}      & \colhead{(2)}    & \colhead{(3)} & \colhead{(4)}          & \colhead{(5)}        & \colhead{(6)}
}
\startdata
1   & \object{J120655.63+501737.1}  & SBS 1204+505B & Apr-07	& 2700 &  1.0   \\
2   & \object{J121600.04+124114.3}  & Mrk 764       & May-07    & 3600 &  0.7 \\
3   & \object{J121607.08+504930.0}  & Mrk 1469      & Apr-07	& 1700 &  0.9  \\
4   & \object{J141238.14+391836.5}  & NGC 5515      & May-07    & 3840 &  0.7  \\
5   & \object{J143031.18+524225.8}  & SBS 1428+529  & Apr-07	& 2400 &  0.9  \\
6   & \object{J144049.35+505009.2}  & SBS 1439+510  & Apr-07	& 3600 &  0.9 \\
7   & \object{J153810.05+573613.1}  & SBS 1537+577  & Apr-07	& 4800 &  1.2   \\
8   & \object{J162952.88+242638.3}  & Mrk883        & Apr-07	& 2400 &  0.8 \\
9   & \object{J212851.19--010412.4} & IC 1385       & Oct-06	& 3900 &  1.0 \\
10 & \object{J234428.81+134946.0}   & ...           & Oct-06	& 3600 &  0.8 
\enddata
\tablecomments{Col. 1: galaxy identification. Col. 2: SDSS ID. Col. 3: other name. Col. 4: observation dates. Col. 5: exposure time. Col. 6: average seeing.}
\end{deluxetable}

\begin{deluxetable}{cccccccc}
\tabletypesize{\scriptsize}
\tablecolumns{8}
\tablewidth{0pc}
\tablecaption{{Host Galaxy Properties {\bf FROM SBD DECOMPOSITION}}\label{tab:tabla2}}
\tablehead{\colhead{Galaxy} & \colhead{$n$} & \colhead{$r_e$}  &  \colhead{($b/a)_{bulge}$}  & \colhead{($PA)_{bulge}$} & \colhead{$\chi^2$} & \colhead{Morphological} & \colhead{Classical Bulge\,/\,}\\ 
\colhead{\#}   &       & \colhead{$h$}    & \colhead{($b/a)_{disk}$}     & \colhead{($PA)_{disk}$}    &          &\colhead{Type}                   & \colhead{Pseudobulge}\\
            &             & \colhead{$a_{bar}$}& \colhead{($b/a)_{bar}$} &  \colhead{($PA)_{bar}$} &    &    & \\
\colhead{(1)} & \colhead{(2)} & \colhead{(3)}    &  \colhead{(4)}    & \colhead{(5)}   & \colhead{(6)}   & \colhead{(7)} & \colhead{(8)}
}                            
\startdata
1      & 2.8 $\pm$ 1.2   & 2.1 $\pm$ 0.9 & 0.79 $\pm$ 0.02 & 175.9$\pm$ 1.6 & 7.8 & SBb $^{\ast\ast}$& CB/PB  \\
       &                 & 2.8 $\pm$ 0.7 & 0.59 $\pm$ 0.02 & 49.6 $\pm$ 4.7 &     &   &        \\
       &                 & 5.4 $\pm$ 1.6 & 0.55 $\pm$ 0.02 & 0.6 $\pm$ 4.2  &     &   &        \\
2      & 1.8 $\pm$ 0.6   & 4.8 $\pm$ 1.1 & 0.95 $\pm$ 0.02 & 2.3 $\pm$ 3.3  & 4.9 & Sb(r)I$^{\ast}$& CB{\bf /PB}  \\
       &                 & 16.3 $\pm$ 1.0& 0.78 $\pm$ 0.01 & 55.6 $\pm$ 5.8 &     &   &        \\ 
       &                 &\nodata           & \nodata            &\nodata          &     &   &        \\
3      & 1.6 $\pm$ 0.5   & 1.7 $\pm$ 0.4 & 0.34 $\pm$ 0.03 & 78.5 $\pm$ 2.7 & 3.6 &Sb $^{\ast\ast}$& PB   \\
       &                 & 3.4 $\pm$ 0.3 & 0.34 $\pm$ 0.01 & 68.8 $\pm$ 4.2 &     &   &        \\
       &                 &\nodata           & \nodata            &\nodata          &     &   &       \\
4      & 1.4 $\pm$ 0.8   & 0.9 $\pm$ 0.5 & 0.62 $\pm$ 0.07 & 102.8 $\pm$ 1.5 & 4.1 & Sab $^{\ast}$ & PB     \\
       &                 & 3.2 $\pm$ 0.2 & 0.48 $\pm$ 0.02 & 104.3 $\pm$ 2.3 &     &   &        \\
       &                 &\nodata           & \nodata            &\nodata           &     &   &        \\
5      & 1.3 $\pm$ 0.6   & 0.6 $\pm$ 0.5 & 0.64 $\pm$ 0.01 & 168.4 $\pm$ 2.9 & 2.1 & SBa-SBab $^{\ast\ast}$ & PB    \\
       &                 & 4.7 $\pm$ 0.7 & 0.63 $\pm$ 0.02 & 154.4 $\pm$ 4.9 &     &   &        \\
       &                 & 7.0 $\pm$ 1.3 & 0.37 $\pm$ 0.02 & 164.4 $\pm$ 4.7 &     &   &       \\
6      & 1.6 $\pm$ 0.8   & 1.9 $\pm$ 0.9 & 0.49 $\pm$ 0.08 & 124.7 $\pm$ 2.0 & 3.5 & SBa$^{\ast\ast}$  & CB/PB  \\
       &                 & 2.8 $\pm$ 1.4 & 0.57 $\pm$ 0.03 & 40.3 $\pm$ 5.1  &     &   &        \\
       &                 & 2.8 $\pm$ 3.8 & 0.45 $\pm$ 0.03 & 120.2 $\pm$ 3.5 &     &   &        \\
7      & 1.6 $\pm$ 0.3   & 1.9 $\pm$ 0.5 & 0.84 $\pm$ 0.02 & 2.3 $\pm$ 3.6   & 5.0 &Sc$^{\ast\ast}$& PB  \\
       &                 & 5.4 $\pm$ 0.7 & 0.76 $\pm$ 0.01 & 105.4 $\pm$ 3.5 &     &   &        \\
       &                 & \nodata            & \nodata            & \nodata           &     &   &       \\
8      & 1.8 $\pm$ 0.3   & 1.2 $\pm$ 0.4 & 0.83 $\pm$ 0.02 & 31.5 $\pm$ 3.2  & 6.4 & peculiar & uncertain   \\
       &                 & 5.9 $\pm$ 0.3 & 0.71 $\pm$ 0.01 & 62.3 $\pm$ 2.9 &     &   &       \\
       &                 & \nodata           & \nodata            & \nodata           &     &   &      \\
9      &\nodata          & \nodata           & \nodata             & \nodata           & \nodata & Sab$^{\ast\ast}$ & uncertain    \\
       &                 & 3.3 $\pm$ 0.9 & 0.91 $\pm$ 0.02 & 135.2 $\pm$ 6.3 &     &   &        \\
       &                 & \nodata          & \nodata            &\nodata           &     &   &        \\
10     &\nodata          & 0.6 $\pm$ 0.5 & 0.80 $\pm$ 0.02 & 118.2 $\pm$ 3.1 & 3.1 & SBab$^{\ast\ast}$  & uncertain   \\
       &                 & 5.2 $\pm$ 0.9 & 0.65 $\pm$ 0.02 & 124.8 $\pm$ 5.8 &     &   &        \\
       &                 & 9.9 $\pm$ 1.5 & 0.30 $\pm$ 0.02 & 133.4 $\pm$ 3.7 &     &   &      
\enddata
\tablecomments{Parameters of the host galaxy derived from the 
{\bf surface brightness decomposition,} SBD. 
Col.  1: galaxy  \#  (cf. Table~\ref{tab:tabla1}).   Col. 2:  S\'ersic
index.  Col.  3:  bulge   effective  radius  in  kpc  ($r_{e}$),  disk
scale-length  in kpc  ($h$), bar  length ($a_{bar}$)  in kpc.  Col. 4:
bulge, disk,  and bar  axis ratio ($b/a)_{bulge}$,  ($b/a)_{disk}$ and
($b/a)_{bar}$,  respectively. Col.  5: bulge,  disk, and  bar position
angle     ($PA)_{bulge}$,      ($PA)_{disk}$     and     ($PA)_{bar}$,
respectively. Col. 6: $\chi^{2}$ of the fit.  Col. 7: morphology given
by NED ($^{\ast}$) and derived  in this work ($^{\ast\ast}$); galaxy 7
is peculiar,  probably a merger. Col.  8: host bulge  is most probably
classical (CB), pseudobulge (PB), or a combination CB/PB.}
\end{deluxetable}

\begin{deluxetable}{cccccc}
\tabletypesize{\scriptsize}
\tablecolumns{6}
\tablewidth{0pc}
\tablecaption{Black Hole Mass Estimates \label{tab:tabla3}}
\tablehead{
\colhead{Galaxy} & \colhead{Morphological} & \colhead{Bulge} & \colhead{$\log M_{\rm BH}^{\sigma_{\ast}}$} & \colhead{$\log M_{\rm BH}^{\sigma_{\ast}}$} & \colhead{$\log M_{\rm BH}$} \\
                 & \colhead{Type}  & \colhead{Type}  &  \colhead{classical bulge}  & \colhead{pseudobulge}  & \\
\colhead{\#}     &                 &                 &  \colhead{(M$_{\odot})$}     & \colhead{(M$_{\odot})$} & \colhead{(M$_{\odot})$}    \\
\colhead{(1)} & \colhead{(2)} & \colhead{(3)} & \colhead{(4)} & \colhead{(5)} & \colhead{(6)} 
}
\startdata
1   & SBb      & CB/PB*      & {\bf 6.54$\pm$0.04} & {\bf 5.95$\pm$0.15} & {\bf 7.08$\pm$0.14}        \\
2   & Sb(r)    & CB{\bf /PB} & {\bf 6.97$\pm$0.06} & {\bf 6.35$\pm$0.05} & {\bf 7.48$\pm$0.16}       \\
3   & Sb       & PB*         &                     & {\bf 6.90$\pm$0.09} & {\bf 7.81$\pm$0.14} \\
    &          &             &                     &                     &                           \\
4   & Sab      & PB*         &                     & {\bf 7.45$\pm$0.25} & {\bf 7.01$\pm$0.18}                 \\
5   & SBa-SBab & PB*         &                     & {\bf 7.13$\pm$0.18} & {\bf 6.87$\pm$0.25}                \\
6   & SBa      & CB/PB       & {\bf 8.09$\pm$0.24} & {\bf 7.38$\pm$0.32} & {\bf 7.04$\pm$0.23}         \\
7   & Sc       & PB          &                     & {\bf 5.96$\pm$0.14} & {\bf 7.13$\pm$0.10}   \\
    &          &             &                     &                     &                        \\
8   & Peculiar & unknown*    & {\bf 8.07$\pm$0.15} & {\bf 7.36$\pm$0.24} & {\bf 7.28$\pm$0.11}                  \\
9   & Sab      & unknown     & {\bf 6.26$\pm$0.04} & {\bf 5.69$\pm$0.21} & {\bf 6.82$\pm$0.08}  \\
    &          &             &                     &                     &                               \\
10  & SBab     & unknown     & {\bf 6.60$\pm$0.17} & {\bf 6.00$\pm$0.02} & {\bf 6.54$\pm$0.16} \\
    &          &             &                     &                     &                             
\enddata
\tablecomments{Col.   1:  galaxy   \#   (cf.  Table~\ref{tab:tabla1}).
  Col. 2: morphological classification.  Col. 3: classical bulge (CB),
  pseudobulge  (PB).  Asterisk  indicates objects  that  display
    double-peaked  emission  lines  (Paper  I).  Col.  4:  $M_{\rm BH}$
  estimates    when    the   host    harbors    a   classical    bulge
  \citep[see][]{2005SSRv..116..523F}. Col. 5: {BH} mass estimates when
  the           host          harbors           a          pseudobulge
  \citep[see][]{2008MNRAS.386.2242H}.  There are  3 objects  with bulges
    showing    both    possibilities    (classical
  bulge/pseudobulge), thus   the  $M_{\rm BH}$   mass  was   estimated   using  both
  methods.  Col.  6:  {\bf BH  mass  estimates using the scaling relations given by 
\citet{2006ApJ...641..689V} derived in Paper I}. 
   }
\end{deluxetable}

\end{document}